\documentclass[conference]{IEEEtran}
\IEEEoverridecommandlockouts

\usepackage[T1]{fontenc}
\usepackage[utf8]{inputenc}
\usepackage{cite}
\usepackage{xurl}
\usepackage{hyperref}
\usepackage{booktabs}
\usepackage{array}
\usepackage{tabularx}
\usepackage{enumitem}
\usepackage{amsmath}
\usepackage{graphicx}
\usepackage{xcolor}
\usepackage{tikz}
\usetikzlibrary{arrows.meta,positioning,fit,shapes.misc,calc}

\hypersetup{hidelinks}
\setlist[itemize]{leftmargin=*,noitemsep,topsep=2pt}
\setlist[enumerate]{leftmargin=*,noitemsep,topsep=2pt}
\newcolumntype{Y}{>{\raggedright\arraybackslash}X}
\newcommand{\riskterm}[1]{\textit{#1}}

\title{Agentic AI and the Industrialization of Cyber Offense:\\
Forecast, Consequences, and Defensive Priorities for Enterprises and the Mittelstand}

\author{\IEEEauthorblockN{Christopher Koch}
\IEEEauthorblockA{Independent Researcher}}

\begin{document}
\maketitle

\begin{abstract}
Agentic AI systems can plan, call tools, inspect code, interact with web applications, and coordinate multi-step workflows. These same capabilities change the economics of cyber offense. The central near-term risk is not that every low-skill criminal immediately becomes a frontier exploit researcher; it is that agentic AI compresses the attack lifecycle by lowering the cost of reconnaissance, phishing, credential abuse, vulnerability triage, exploit adaptation, and post-compromise decision support. This paper synthesizes current public evidence from national cyber-security agencies, industry threat reports, agent-security guidance, and research on LLM agents' cyber capabilities. It introduces a Three-Channel Agentic Cyber-Risk Model and an Agentic Attack Compression Model, uses the 2026 Linux kernel ``Copy Fail'' incident as a case study for foothold-to-root acceleration, and develops a 2026--2028 forecast for large enterprises and the German/European Mittelstand. The paper concludes with a prioritized defense roadmap. Organizations should treat agentic AI security as an immediate operational problem: identity, phishing-resistant authentication, patch velocity, CI/CD and Linux/container hardening, agent governance, telemetry, and recovery readiness must be strengthened now.
\end{abstract}


\section{Introduction}
Agentic AI marks a shift from content generation to goal-directed action. Modern agents can reason over instructions, retrieve data, call APIs, write code, run tools, and iterate over intermediate results. In legitimate enterprise settings, these features promise productivity gains in software engineering, operations, customer service, analytics, and security operations. The same structure is dual-use: an agent that can research, plan, generate code, test hypotheses, and automate tool use can also help an attacker scale phishing, reconnaissance, vulnerability triage, credential abuse, and post-compromise operations.

This paper asks a practical question: \textit{if agentic AI continues to improve, how should enterprises and the Mittelstand expect the cyber threat landscape to change, and what should they do now?} The paper is defensive and strategic. It does not provide exploit instructions, payloads, or operational attack steps. Instead, it synthesizes public evidence and translates it into security planning.

The key finding is that agentic AI should be understood as an \riskterm{attack compression technology}. It reduces the time, skill, and cost required to move through phases of an attack. This is already visible in national assessments and threat reporting. The UK National Cyber Security Centre (NCSC) assesses that AI will almost certainly continue to make cyber intrusion operations more effective and efficient, increasing the frequency and intensity of cyber threats through 2027 \cite{ncsc2027}. Microsoft describes AI-automated phishing and faster exploitation of known security gaps as part of the 2025 threat landscape \cite{microsoft_mddr}. Verizon's 2025 DBIR reports that exploitation of vulnerabilities increased by 34\%, while credential abuse and vulnerability exploitation remained leading initial access vectors \cite{verizon_dbir}. These trends matter more as agents become easier to run, chain, and commercialize.

The recent Linux kernel ``Copy Fail'' vulnerability, CVE-2026-31431, illustrates the operational importance of this shift. It is a local privilege-escalation flaw: not remotely exploitable by itself, but highly consequential once an attacker has any local code execution, for example through compromised credentials, a malicious CI job, or a container foothold. Microsoft emphasizes its impact in cloud, CI/CD, and Kubernetes environments \cite{ms_copyfail}; Ubuntu rated it high severity and published mitigations and fixes \cite{ubuntu_copyfail}; NVD records it in CISA's Known Exploited Vulnerabilities catalog with a May 15, 2026 remediation due date \cite{nvd_copyfail}. The lesson is broader than one CVE: when initial access becomes cheaper and privilege escalation becomes easier to operationalize, defenders must assume that the time from foothold to impact will continue to shrink.

\subsection{Contributions}
This paper makes four contributions.
\begin{enumerate}
    \item It proposes a \textit{Three-Channel Agentic Cyber-Risk Model}: attacker-side augmentation, agentic systems as targets, and internal autonomous agents as risky actors.
    \item It proposes the \textit{Agentic Attack Compression Model} (AACM), explaining how agentic AI lowers attacker cost across the intrusion lifecycle without assuming full autonomous super-hacking.
    \item It provides a 2026--2028 forecast for enterprises and the Mittelstand, distinguishing high-confidence near-term effects from more speculative frontier-agent effects.
    \item It presents a prioritized defense roadmap across identity, email trust, patch velocity, Linux/container/CI/CD hardening, agent governance, SOC telemetry, and recovery.
\end{enumerate}

\section{Method and Scope}
This paper is a bounded evidence synthesis rather than an empirical benchmark. Sources were selected from four categories: (i) official government and agency assessments; (ii) industry threat reports with large telemetry or incident-response vantage points; (iii) peer-reviewed or preprint research on LLM/agent cyber capability; and (iv) vulnerability advisories and vendor analyses for the Copy Fail case. The goal is not to average incompatible metrics, but to identify converging evidence chains.

The synthesis prioritizes sources that meet at least one of the following criteria: (1) they describe macro-level cyber threat trends; (2) they measure or demonstrate agentic cyber capability; (3) they document agentic AI security risks and mitigations; (4) they provide actionable incident or vulnerability evidence relevant to enterprise defense. The paper excludes procedural offensive content and intentionally avoids details that would meaningfully enable exploitation.

\section{Background: What Agentic AI Changes}
Traditional generative AI primarily outputs text, code, or media in response to a prompt. Agentic AI adds three capabilities that are especially security-relevant:
\begin{itemize}
    \item \textbf{Tool use}: agents can invoke scanners, browsers, shells, APIs, ticketing systems, repositories, and cloud services.
    \item \textbf{Stateful planning}: agents can decompose objectives, maintain memory, retry strategies, and adapt over multiple steps.
    \item \textbf{Autonomous action}: agents can execute actions that affect systems, identities, data, or business processes.
\end{itemize}

Security guidance increasingly treats this autonomy as a new risk boundary. A 2026 joint cybersecurity information sheet on careful adoption of agentic AI highlights privilege, design and configuration, behavior, structural, accountability, and supply-chain risks, and recommends incremental adoption, continuous assessment, monitoring, governance, accountability, and human oversight \cite{nsa_agentic}. NIST's AI Agent Standards Initiative similarly frames secure, interoperable agents as a standardization priority \cite{nist_agents}. OWASP's agentic AI guidance frames agentic systems as a distinct security problem because they combine LLM risks with action, tool use, and autonomy \cite{owasp_agentic}. The OWASP Top 10 for LLM Applications also lists ``Excessive Agency'' as a risk caused by excessive functionality, permissions, or autonomy \cite{owasp_llm_top10}.

For cyber offense, the practical implication is not only better text generation. Agentic systems can support entire workflows: gather information, translate it into target-specific hypotheses, write or adapt scripts, summarize documentation, generate convincing lures, triage errors, and route tasks to subagents. Research has demonstrated that LLM agents can autonomously hack vulnerable websites in controlled settings \cite{fang_web}, exploit one-day vulnerabilities when given CVE descriptions \cite{fang_oneday}, and improve zero-day exploitation performance with teams of agents \cite{fang_zeroday}. Larger-scale benchmarks such as CyberGym show that real-world vulnerability reproduction remains difficult, but also that agents can produce direct security impact and discover new vulnerabilities in some settings \cite{cybergym}. The correct conclusion is therefore balanced: agentic AI is not omnipotent, but it is already strong enough to change attacker economics.

\section{Three-Channel Agentic Cyber-Risk Model}
Agentic AI introduces cyber risk through three channels, summarized in Table~\ref{tab:channels}. Many discussions focus only on malicious users asking models for harmful output. That is too narrow. Companies must also secure the agentic systems they deploy and the autonomous actions those systems may take inside the enterprise.

\begin{table*}[t]
\centering
\caption{Three-channel model of agentic AI cyber risk}
\label{tab:channels}
\begin{tabularx}{\textwidth}{p{2.9cm}Y Y Y}
\toprule
\textbf{Risk channel} & \textbf{Mechanism} & \textbf{Typical consequence} & \textbf{Primary controls} \\
\midrule
Attacker-side augmentation & Criminals use agents to accelerate reconnaissance, phishing, exploit adaptation, and post-compromise decisions. & More attacks, better personalization, shorter patch windows, lower skill threshold for repeatable campaigns. & Phishing-resistant identity, exposure management, patch velocity, detection engineering, resilience. \\
Agentic systems as targets & Attackers compromise agents, connectors, credentials, memory, tools, or non-human identities. & Data exfiltration, unauthorized tool use, persistence through agent workflows, supply-chain compromise. & Agent inventory, least privilege, secrets management, runtime monitoring, connector governance. \\
Internal agents as risky actors & Legitimate agents act with excessive autonomy, ambiguous goals, weak guardrails, or poor approvals. & Unintended state changes, data leakage, unsafe automation, compliance failures. & Human approval gates, bounded tools, policy-as-code, test harnesses, audit traces, rollback. \\
\bottomrule
\end{tabularx}
\end{table*}

This model is useful because it separates three defensive obligations. First, companies must defend against adversaries using agentic AI. Second, companies must secure their own agent infrastructure. Third, companies must prevent legitimate agents from becoming internal sources of uncontrolled action. The 2026 joint agentic AI guidance emphasizes that securing agentic systems requires established cybersecurity principles adapted to autonomy, interconnection, and evolving capabilities \cite{nsa_agentic}. Microsoft similarly notes that autonomous systems can invoke tools, call APIs, access data, and coordinate across services, expanding the attack surface and risk of unintended actions or exfiltration \cite{microsoft_agentic_risk}.

\section{The Agentic Attack Compression Model}
\begin{figure}[t]
\centering
\begin{tikzpicture}[font=\scriptsize, node distance=3.6pt, every node/.style={align=center}]
\tikzstyle{box}=[draw, rounded corners, minimum height=0.55cm, text width=2.95cm, fill=gray!10]
\node[box] (r) {Reconnaissance\\OSINT summarization};
\node[box, below=of r] (p) {Phishing and impersonation\\localized, personalized lures};
\node[box, below=of p] (c) {Credential abuse\\session, helpdesk, SaaS workflows};
\node[box, below=of c] (v) {Vulnerability matching\\known CVEs, patch gaps};
\node[box, below=of v] (e) {Exploit adaptation\\debug, test, iterate};
\node[box, below=of e] (x) {Post-compromise planning\\persistence, lateral movement};
\draw[-{Latex}] (r) -- (p);
\draw[-{Latex}] (p) -- (c);
\draw[-{Latex}] (c) -- (v);
\draw[-{Latex}] (v) -- (e);
\draw[-{Latex}] (e) -- (x);
\node[draw, dashed, rounded corners, fit=(r)(p)(c)(v)(e)(x), inner sep=4pt, label={[align=center]above:Agentic Attack Compression}] {};
\end{tikzpicture}
\caption{Agentic Attack Compression Model. Agentic AI does not need to invent novel attack classes to increase risk. It compresses existing phases by lowering cost, increasing speed, and reducing coordination barriers.}
\label{fig:aacm}
\end{figure}
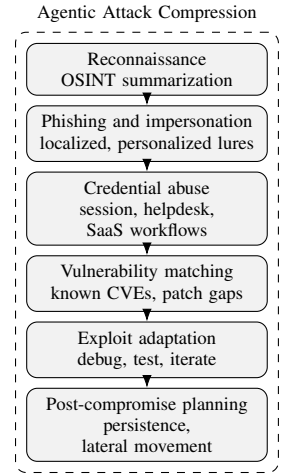

We define \textit{attack compression} as the reduction of attacker time, skill, cost, or coordination effort required to complete one or more steps of an intrusion. Agentic AI compresses attacks through six mechanisms, shown in Fig.~\ref{fig:aacm}.

\subsection{Reconnaissance and Target Selection}
Agents can summarize public websites, job postings, cloud documentation, GitHub repositories, exposed services, leaked credentials, and supplier relationships. This makes reconnaissance cheaper and more scalable. The risk for SMEs and Mittelstand companies is that they no longer need to be individually high-value targets; they can be swept into automated campaigns that generate credible target-specific pretexts.

\subsection{Phishing, Business Email Compromise, and Impersonation}
Phishing has always been high-return. AI improves language quality, localization, personalization, and campaign variation. Microsoft reports AI-automated phishing and multi-stage attack chains in its 2025 Digital Defense Report \cite{microsoft_mddr}. OpenAI's misuse reporting also observes threat actors using AI in combination with traditional assets such as websites and social media accounts, often bolting AI onto older playbooks rather than obtaining fundamentally novel capabilities \cite{openai_misuse}. This distinction is important: the immediate danger is scale and credibility, not necessarily autonomous end-to-end compromise.

\subsection{Credential Abuse and Identity-Centered Intrusion}
Modern attacks increasingly target identity and cloud control planes rather than malware alone. Verizon reports credential abuse and vulnerability exploitation as leading initial access vectors \cite{verizon_dbir}; CrowdStrike reports that 81\% of hands-on-keyboard intrusions were malware-free and highlights helpdesk impersonation, vishing, SaaS, and cloud lateral movement \cite{crowdstrike_hunting}. Agentic AI reinforces this trend because identity workflows are natural language and process-heavy: helpdesk scripts, reset procedures, approval messages, and social validation can all be assisted by AI.

\subsection{Vulnerability Exploitation and Patch-Window Compression}
NCSC judges that AI-enabled tools will almost certainly enhance threat actors' ability to exploit known vulnerabilities and reduce the time between disclosure and exploitation \cite{ncsc2027}. Academic work on one-day vulnerability exploitation supports this concern: when supplied with CVE descriptions, a frontier LLM agent exploited a substantial portion of a small benchmark of real-world vulnerabilities, while other tested models and scanners did not \cite{fang_oneday}. The implication for defenders is simple: published advisories become more actionable for attackers, and patch latency becomes a primary business risk metric.

\subsection{Post-Compromise Operations}
Once an attacker has access, agents can help summarize internal files, identify privilege paths, draft command sequences, and prioritize next steps. This is especially relevant for smaller teams of criminals because it reduces the need for deep, continuous expertise at every stage. However, advanced multi-step exploitation still depends on access, context, tooling, and operator judgment. The forecast should therefore avoid both complacency and exaggeration.

\section{Case Study: Copy Fail and the Foothold-to-Root Problem}
CVE-2026-31431, known as Copy Fail, is a Linux kernel local privilege-escalation vulnerability. Public advisories describe it as affecting mainstream Linux distributions with vulnerable kernels and enabling a local, low-privilege user to escalate to root. Microsoft highlights that it is not remotely exploitable in isolation, but becomes highly impactful when chained with initial access such as SSH access, malicious CI job execution, or container footholds; it also notes risk in cloud, CI/CD, and Kubernetes environments \cite{ms_copyfail}. Ubuntu rated it high severity and released mitigation guidance disabling the affected kernel module pending kernel updates \cite{ubuntu_copyfail}. CERT-EU recommended prioritizing Kubernetes nodes and CI/CD runners exposed to untrusted workloads \cite{cert_eu_copyfail}. NVD records the CVE in CISA's Known Exploited Vulnerabilities catalog with a due date of May 15, 2026 \cite{nvd_copyfail}.

The case matters for agentic AI for three reasons. First, AI-assisted code analysis can reduce the cost of finding deep software flaws, even if expert validation and responsible disclosure remain essential. Second, public exploit availability lets less sophisticated attackers benefit from sophisticated discovery work. Third, local privilege escalation changes the business meaning of initial access: a compromised developer account, CI runner, web application process, or container is not a limited nuisance if it can become root on unpatched systems.

The Copy Fail example therefore supports a broader defensive principle: organizations should prioritize not only vulnerability severity, but also \textit{chainability}. A local privilege escalation may be business-critical when it is easy to combine with phishing, credential theft, CI/CD abuse, or container compromise.

\section{Forecast: 2026--2028}
Table~\ref{tab:forecast} summarizes the expected development path. The forecast follows the evidence pattern in NCSC, Microsoft, Verizon, OpenAI, CrowdStrike, OWASP, and academic agent studies: existing tactics become faster, more scalable, and easier to operationalize; only some actors will exploit frontier capabilities early; and defenders can also use AI to improve detection, triage, and remediation.

\begin{table*}[t]
\centering
\caption{Forecast for agentic AI effects on cyber risk}
\label{tab:forecast}
\begin{tabularx}{\textwidth}{p{2.2cm}Y Y Y}
\toprule
\textbf{Horizon} & \textbf{Likely attacker uplift} & \textbf{Affected companies} & \textbf{Defensive implication} \\
\midrule
0--6 months & More convincing phishing, faster credential abuse, faster weaponization of public advisories, more AI-supported script adaptation. & All organizations; SMEs and Mittelstand exposed through email, SaaS, suppliers, and hosted Linux systems. & Prioritize identity, email trust, urgent patching, external attack surface, backup/restore, and Linux/container emergency response. \\
6--18 months & Semi-autonomous recon, exploit-chain assistance, SaaS/cloud workflow abuse, better helpdesk impersonation, targeting of AI agents and non-human identities. & Enterprises with many SaaS/cloud identities; Mittelstand firms dependent on managed IT and exposed remote services. & Move to phishing-resistant MFA, govern non-human identities, harden CI/CD, monitor cloud control planes, and inventory agents/tools. \\
18--36 months & Crimeware-as-a-service packages agentic workflows; more attackers can run multi-step campaigns; capable actors use AI for vulnerability research and exploit development. & High-value corporates, critical suppliers, software firms, MSPs, industrial firms, and mid-market manufacturers. & Build AI-aware SOC workflows, threat-informed exposure management, agent runtime controls, continuous validation, and cyber resilience exercises. \\
\bottomrule
\end{tabularx}
\end{table*}

\subsection{Will Every Small Criminal Become an Advanced Hacker?}
The answer is no, but the question understates the risk. The expected change is not universal elite capability; it is capability \textit{packaging}. Many criminals will not discover novel kernel bugs, but they may use AI-enhanced services that generate phishing kits, adapt public exploits, automate reconnaissance, and guide post-compromise actions. OpenAI reports that malicious actors often use AI to move faster within existing playbooks rather than gain novel offensive capability from models alone \cite{openai_misuse}. NCSC similarly assesses that only highly capable actors are likely to harness the full potential of frontier AI in advanced operations in the near term, while a broader set of actors repurpose commercial and open-source models to uplift capability \cite{ncsc2027}. This is still dangerous because cybercrime scales through services, reuse, and automation.

\subsection{Expected Impact for Enterprises}
Large enterprises face higher absolute attack volume and more complex blast radius. They usually have better tooling, but they also have more identities, vendors, SaaS applications, cloud tenants, CI/CD pipelines, data stores, and unmanaged workflows. Agentic AI increases the value of weak processes: helpdesk reset procedures, approval workflows, exception handling, vendor onboarding, and shadow IT. It also creates new internal attack surfaces when companies deploy their own agents with excessive permissions.

\subsection{Expected Impact for the Mittelstand}
The Mittelstand is exposed differently. Smaller and mid-sized firms often lack 24/7 SOC coverage, mature vulnerability management, identity governance, red-team testing, and dedicated cloud security staff. ENISA notes that SMEs represent 99\% of EU businesses and that, in surveyed SMEs, 90\% said cybersecurity issues would seriously affect their business within a week, with 57\% saying such issues could likely lead to bankruptcy or going out of business \cite{enisa_sme}. Allianz ranks cyber incidents as the top global business risk for 2026 and notes that smaller and mid-sized enterprises are increasingly targeted and under pressure due to limited security resources \cite{allianz2026}. For the Mittelstand, agentic AI therefore raises the probability that attackers can economically target companies that previously appeared too small to justify tailored operations.

\section{Consequences for Corporate Security Strategy}
Agentic AI changes security planning in five ways.

\subsection{Patch Latency Becomes a Board Metric}
The Copy Fail case and NCSC forecast point in the same direction: defenders have less time between disclosure, exploit availability, and exploitation. Patch latency should be tracked for internet-facing systems, endpoints, servers, cloud workloads, Kubernetes nodes, CI/CD runners, VPNs, identity providers, and managed appliances. CISA KEV membership, exploit availability, exposure, and chainability should override generic CVSS-only prioritization.

\subsection{Identity Is the Primary Attack Surface}
Agentic phishing, vishing, helpdesk impersonation, and SaaS abuse make identity controls central. Phishing-resistant MFA, conditional access, device posture, privileged access management, and non-human identity governance become baseline controls, not maturity luxuries.

\subsection{CI/CD and Linux Hosts Are High-Value Targets}
Build runners, container nodes, developer workstations, artifact registries, and Kubernetes clusters combine credentials, code, secrets, and privileged execution. Copy Fail illustrates why a low-privilege execution context on a Linux host can become a full compromise if kernel patching and containment lag.

\subsection{Agents Themselves Become Enterprise Assets}
As companies deploy internal agents, attackers may target agent credentials, tool permissions, memory, plugins, connectors, and approval workflows. The joint agentic AI guidance and OWASP agentic risk taxonomy both imply that agents should be inventoried, scoped, monitored, and governed like high-impact digital workers rather than treated as simple chatbots \cite{nsa_agentic,owasp_agentic}. AWS and Microsoft guidance likewise map agentic risks to concrete controls such as scoping, threat modeling, and monitoring \cite{aws_agentic,microsoft_agentic_risk}.

\subsection{Resilience Matters More Than Prevention Alone}
Agentic AI may increase attack frequency and speed. No organization can rely on perfect prevention. Tested backups, restoration exercises, incident-response playbooks, segmentation, crisis communications, and supplier failover are therefore part of AI-era security rather than afterthoughts.

\section{What Companies Should Do Now}
Table~\ref{tab:roadmap} provides a prioritized roadmap. It is deliberately pragmatic: the first actions reduce the highest-probability losses without requiring a complete security transformation.

\begin{table*}[t]
\centering
\caption{Defensive roadmap for the agentic AI cyber-risk era}
\label{tab:roadmap}
\begin{tabularx}{\textwidth}{p{2.1cm}p{3.0cm}Y}
\toprule
\textbf{Timeframe} & \textbf{Priority area} & \textbf{Actions} \\
\midrule
0--30 days & Identity and email & Enforce phishing-resistant MFA for admins and remote access; review helpdesk reset procedures; enable/validate SPF, DKIM, and DMARC; create out-of-band approval rules for payments, bank changes, and privileged account changes. \\
0--30 days & Copy Fail / Linux exposure & Inventory Linux kernels, Kubernetes nodes, CI runners, and shared hosting systems; apply vendor patches or mitigations; prioritize systems with untrusted workloads or local code execution; add detection for suspicious privilege-escalation behavior. \\
0--30 days & Backups and recovery & Verify immutable/offline backups for critical systems; run at least one restore test; identify minimum viable operations if identity, email, ERP, or file storage is unavailable. \\
30--90 days & Patch velocity & Implement KEV- and exploit-aware prioritization; track patch latency as a metric; include cloud appliances, VPNs, identity providers, web apps, and container hosts; define emergency change windows. \\
30--90 days & CI/CD and cloud & Use ephemeral or isolated runners where feasible; reduce privileged containers; protect secrets; enforce least privilege for build identities; monitor cloud control-plane activity and anomalous access. \\
30--90 days & SOC and telemetry & Integrate identity, endpoint, cloud, email, and network signals; detect impossible travel, MFA fatigue, suspicious OAuth grants, abnormal admin actions, and rapid privilege escalation; use AI defensively for triage but keep human approval for destructive actions. \\
90--180 days & Agent governance & Inventory internal AI agents and tools; assign owners; restrict tools and data by role; require approval for state-changing actions; log traces securely; test agents against prompt injection, excessive agency, data exfiltration, and tool misuse. \\
90--180 days & Supplier and Mittelstand resilience & Require critical suppliers and MSPs to document MFA, backup, patch, and incident-response posture; rehearse supplier outage scenarios; build mutual aid or MSSP coverage for smaller entities lacking 24/7 security. \\
\bottomrule
\end{tabularx}
\end{table*}

\subsection{Identity and Anti-Phishing Controls}
Organizations should move from ``MFA exists'' to phishing-resistant MFA for privileged and high-risk users. They should also redesign business processes that attackers exploit: payment approvals, vendor bank changes, password resets, external file-sharing, and executive requests. AI makes fraudulent communication more credible; process verification must therefore become more explicit.

\subsection{Vulnerability and Exposure Management}
Vulnerability management should move from periodic scanning to exposure- and exploit-aware triage. A vulnerability should be prioritized when it is externally exposed, present in KEV, has public exploit code, enables privilege escalation, affects shared infrastructure, or chains with realistic initial access. The Copy Fail class of issue should trigger emergency response for CI/CD and container infrastructure even if remote exploitation is not possible by itself.

\subsection{Agentic AI Security Controls}
When companies deploy their own agents, they should apply least privilege, tool allowlists, scoped identities, human approval gates, policy-as-code, trace logging, and isolation. Agents should not inherit broad human accounts. They should receive dedicated non-human identities with bounded permissions and revocable credentials. Their memory and tool outputs should be treated as security-relevant artifacts.

\subsection{Mittelstand-Specific Priorities}
For the Mittelstand, the priority is not to build a hyperscale SOC overnight. It is to close the most common gaps with high leverage: managed endpoint protection, phishing-resistant MFA for privileged accounts, secure backups, patch SLAs for exposed systems, clear MSP contracts, emergency contacts, and incident-response drills. Where internal capacity is limited, shared services, managed detection and response, or sector-level security cooperation may be more realistic than purely internal teams.

\section{Discussion}
The evidence supports a nuanced position. Agentic AI is not magic; current agents still fail in many real-world vulnerability tasks, and advanced exploit development remains difficult. But defenders should not wait for perfect autonomy before acting. Cybercrime rarely requires perfect tools. It requires tools that reduce cost enough to make more campaigns profitable. Agentic AI already does this for phishing, reconnaissance, code assistance, and exploit adaptation. The increasing availability of capable models, agent frameworks, and cybercrime services suggests that the long tail of attackers will receive more capability indirectly, through packaged workflows and services.

The defensive opportunity is also significant. AI can help defenders summarize incidents, prioritize vulnerabilities, generate detection logic, validate configurations, and accelerate response. However, defensive AI works best when the organization already has clean telemetry, asset inventories, identity hygiene, and tested processes. Agentic AI does not replace security maturity; it amplifies both strengths and weaknesses.

\section{Threats to Validity}
This paper has four limitations. First, the evidence base mixes official assessments, industry telemetry, vulnerability advisories, and academic experiments; these sources have different incentives and measurement methods. Second, public reporting lags attacker capability and may understate covert use. Third, research benchmarks may overstate or understate real-world attacker effectiveness depending on environmental constraints, tool access, and model availability. Fourth, the forecast is probabilistic. It should be revisited as frontier model capabilities, agent frameworks, regulation, and attacker tooling evolve.

\section{Conclusion}
Agentic AI changes cyber risk by compressing the attack lifecycle. It makes social engineering more scalable, vulnerability information more actionable, exploit adaptation faster, and post-compromise decision support cheaper. The Copy Fail incident demonstrates why this matters operationally: once attackers obtain a foothold, broadly applicable privilege escalation can turn a limited compromise into root-level impact across Linux-heavy cloud, CI/CD, and container environments.

For companies, the correct response is neither panic nor complacency. The near-term priority is to strengthen the controls that reduce attack profitability and blast radius: phishing-resistant identity, hardened email and approval processes, KEV-aware patching, Linux/container/CI hardening, agent governance, high-quality telemetry, and tested recovery. For the Mittelstand, the challenge is sharper because attackers gain scale faster than small organizations gain security capacity. The organizations that adapt now will not eliminate AI-enabled cyber risk, but they will force attackers back into higher-cost, lower-success paths.

\appendices
\section{Search and Review Protocol}
The synthesis used targeted searches between 2024 and 2026 across official cyber-agency publications, vendor threat reports, agentic AI security guidance, and academic cyber-agent literature. Representative query themes included ``AI cyber threat 2027,'' ``agentic AI security guidance,'' ``LLM agents exploit vulnerabilities,'' ``AI automated phishing threat report,'' ``Copy Fail CVE-2026-31431,'' ``Mittelstand cybersecurity SMEs ransomware,'' and ``OWASP agentic AI risks.'' Sources were retained when they directly supported the paper's core claims about agentic cyber capability, threat economics, vulnerability exploitation, enterprise exposure, or defensive controls.

\section{Minimum Board-Level Metrics}
A board or executive committee should track at least the following metrics in the agentic AI cyber-risk era:
\begin{itemize}
    \item phishing-resistant MFA coverage for privileged and high-risk users;
    \item median and 95th-percentile patch latency for KEV/exploited vulnerabilities;
    \item number of internet-facing assets without owner and patch SLA;
    \item number of privileged non-human identities and stale secrets;
    \item backup restore success rate and time to restore critical services;
    \item CI/CD runners and Kubernetes nodes with untrusted workload exposure;
    \item number of production AI agents, their owners, permissions, tools, and audit status;
    \item mean time to detect and contain identity compromise.
\end{itemize}

\section{Ten-Pass Review Log}
The paper was iteratively reviewed against clarity, novelty, evidence strength, safety, and actionability. Table~\ref{tab:reviewlog} summarizes the main improvement made in each pass.

\begin{table}[h]
\centering
\caption{Condensed review and improvement log}
\label{tab:reviewlog}
\begin{tabularx}{\columnwidth}{p{0.65cm}Y}
\toprule
\textbf{Pass} & \textbf{Primary improvement} \\
\midrule
1 & Removed overclaiming about universal autonomous hacking. \\
2 & Added the three-channel risk model. \\
3 & Sharpened the Agentic Attack Compression Model. \\
4 & Strengthened the Copy Fail case and chainability argument. \\
5 & Added Mittelstand-specific consequences. \\
6 & Tightened forecast confidence and time horizons. \\
7 & Expanded action roadmap with CI/CD and Linux priorities. \\
8 & Added agent governance controls. \\
9 & Improved source hierarchy and reduced vendor-only dependence. \\
10 & Final consistency, formatting, safety, and publication-readiness pass. \\
\bottomrule
\end{tabularx}
\end{table}

\end{document}